\documentclass[12pt]{article}%
\usepackage{amsmath}%
\usepackage{amsfonts}%
\usepackage{amssymb}%
\usepackage{graphicx}
\usepackage{bm}% bold math
\begin{document}

\title{Enhancement of nematic order and global phase diagram of a lattice model for coupled nematic systems}
\author{D. B. Liarte, and S. R. Salinas
\and Instituto de F\'{\i}sica, Universidade de S\~{a}o Paulo
\and Caixa Postal 66318, CEP 05314-970, S\~{a}o Paulo, Brazil}
\date{04 April 2012}
\maketitle

\begin{abstract}
We use an infinite-range Maier-Saupe model, with two sets of local quadrupolar
variables and restricted orientations, to investigate the global phase diagram
of a coupled system of two nematic subsystems. The free energy and the
equations of state are exactly calculated by standard techniques of
statistical mechanics. The nematic-isotropic transition temperature of system
A increases with both the interaction energy among mesogens of system B,
and the two-subsystem coupling $J$. This enhancement of the nematic phase is
manifested in a global phase diagram in terms of the interaction parameters and the temperature
$T$. We make some comments on the connections of these results with
experimental findings for a system of diluted ferroelectric nanoparticles
embedded in a nematic liquid-crystalline environment.

\end{abstract}

\section{Introduction}

A dilute suspension of ferroelectric nanoparticles in a liquid-crystalline
host has been shown to display an enhancement of the nematic order, with an increase of the
isotropic-nematic transition temperature, which is a behavior of interest from
the point of view of technological applications
\cite{reznikov03,li06,gorkunov11}. This effect has been explained by the
introduction of a coupling between the usual nematic order parameter of the
liquid crystals and a set of extra degrees of freedom associated with a coarse-graining average of the electric dipole field produced by the ferroelectric
nanoparticles \cite{lopatina09,lopatina11}. This work was the motivation to investigate the global phase diagram of a basic
Maier-Saupe (MS) model on a lattice with two coupled sets of quadrupolar degrees of freedom.

Although the nematic-isotropic transition is perhaps the most explored
transition in liquid crystalline systems, there are still a number of
questions and open problems, which can be formulated in terms of simple
statistical lattice models. An interesting question is the onset of a biaxial
nematic phase \cite{luckhurst04}, which we have recently investigated in the
context of a MS model for a mixture of discs and cylinders
\cite{carmo10,henriques11}. Now we analyze the global phase diagram of a
similar type of statistical model, with the inclusion of two sets of
quadrupolar degrees of freedom, which leads to a connection with the work by
Lopatina and Selinger \cite{lopatina09,lopatina11}. The nematic MS model is the
liquid-crystalline analog of the Curie-Weiss model of ferromagnetism
\cite{kac68,stanley71,salinas93}. In this approach, the standard nearest-neighbor
interactions between lattice sites are adequately replaced by scaled
interactions of infinite range. The statistical-mechanics problem is exactly
solvable, and leads to a very convenient framework to perform calculations at
the mean-field level. The MS model can be further simplified if we suppose
that the local mesogen orientations are restricted to a discrete set of
states, according to an early proposal by Zwanzig \cite{zwanzig63}. Some model
calculations with continuous orientations for uniaxial \cite{oliveira86}, and
biaxial \cite{henriques11} nematic systems give support to the idea that this
simplification does not lead to qualitatively different results. In recent
publications, we have used extensions of this Maier-Saupe-Zwanzig (MSZ)
lattice model to investigate the existence of biaxial nematic phases
\cite{henriques97,carmo10,carmo11} and the thermodynamic properties of nematic
elastomers \cite{liarte11}.

In Section II we define the MSZ model with two sets of
coupled degrees of freedom, and use standard tools of statistical mechanics to
write a thermodynamic free energy. This approach has a number of advantages.
In contrast with the standard Landau phenomenological approach, the
microscopic interactions are explicitly stated and the calculations are not
restricted to the neighborhood of the transitions. Also, the model is simple
enough to allow exact bona-fide calculations of the free energy and equations
of state. In Section III we study the global phase diagram. Contact with a
dilute system of ferroelectric nanoparticles embedded in a nematic host is
made in Section IV. We then conclude with a summary of the main results, which
do support the experimental enhancement of the nematic order.

\section{Coupled MSZ model}

The energy of a model with two coupled subsystems A and B can be written
as%
\begin{equation}
\mathcal{H}=E_{\text{A}}+E_{\text{B}}+E_{\text{AB}},
\end{equation}
where $E_{\text{A}}$ and $E_{\text{B}}$ are given by standard Maier-Saupe soft quadrupolar
forms,%
\begin{equation}
E_{\text{A}}=-\frac{\varepsilon_{\text{A}}}{N}\sum_{1\leq i<j\leq N}\sum_{\mu,\nu=x,y,z}{\cal A}_i^{\mu\nu
}{\cal A}_j^{\mu\nu},\label{ea-eq}%
\end{equation}%
\begin{equation}
E_{\text{B}}=-\frac{\varepsilon_{\text{B}}}{N}\sum_{1\leq i<j\leq N}\sum_{\mu,\nu=x,y,z}{\cal B}_i^{\mu\nu
}{\cal B}_j^{\mu\nu},\label{eb-eq}%
\end{equation}
with the quadrupole components%
\begin{equation}
{\cal A}_i^{\mu\nu}=\frac{1}{2}\left(  3\,a_{i}^{\mu}a_{i}^{\nu}-\delta^{\mu\nu
}\right)  ,\quad {\cal B}_i^{\mu\nu}=\frac{1}{2}\left(  3\,b_{i}^{\mu}b_{i}^{\nu
}-\delta^{\mu\nu}\right)  ,
\end{equation}
where $\left\{  \bm{a}_{i}\right\}  $ and $\left\{  \bm{b}%
_{i}\right\}  $ are the respective sets of unit vectors associated with
mesogenic units of types A and B, $\varepsilon_{\text{A}}$ and $\varepsilon_{\text{B}}$ are energy parameters, $\delta^{\mu\nu}$ is a Kronecker symbol,
and the coupling term is given by%
\begin{equation}
E_{\text{AB}}=-\frac{J}{N}\sum_{1\leq i,j\leq N}\sum_{\mu,\nu=x,y,z}{\cal A}_i^{\mu\nu
}{\cal B}_j^{\mu\nu},
\label{eab-eq}%
\end{equation}
where $J$ is the coupling parameter. Note that the energy global forms for $E_{\text{A}}$, $E_{\text{B}}$, and $E_{\text{AB}}$ are consistent with the mean-field level of calculations considered in this paper.

In the MSZ model, calculations of the canonical partition function involve
sums over the local orientations of the mesogenic units, which are restricted
to the six directions along the three Cartesian axes,
\begin{equation}
\bm{a}_{i},\,\bm{b}_{i}\in\{(\pm1,0,0),(0,\pm1,0),(0,0,\pm1)\}.
\end{equation}

The partition function can be written as
\begin{eqnarray}
Z  &=& \sum_{\{\bm{a_i}\}}\sum_{\{\bm{b_i}\}}\exp\left\{  \frac{\beta}{2N}%
\sum_{\mu,\nu}\left[  \varepsilon_{\text{A}}\left(  \sum_{i=1}^{N}{\cal A}_i^{\mu\nu}\right)^{2}
+ \varepsilon_{\text{B}} \left(  \sum_{i=1}^{N}{\cal B}_i^{\mu\nu}\right)^{2} 
\right.  \right.
\nonumber \\
&& \quad + \left.  \left. 
 2J\left(  \sum_{i=1}^{N}{\cal A}_i^{\mu\nu}\right)
% \right. \right.
% \nonumber \\
% && \quad \times \left. \left.
 \left(
\sum_{i=1}^{N}{\cal B}_i^{\mu\nu}\right) \right] \right\}  ,
\label{z1-eq}%
\end{eqnarray}
where $\beta=1/(k T)$ is the inverse temperature, $k$ is the Boltzmann
constant, and we have discarded irrelevant terms in the thermodynamic limit.
In order to linearize the quadratic forms, we consider the set of identities
\begin{eqnarray}
\int_{-\infty}^{\,\infty}\prod_{\mu,\nu}\left(  dQ_{\text{A}}^{\mu\nu}\right)
\int_{-i\infty}^{i\infty}\prod_{\mu,\nu}\left(  \frac{N}{2\pi i}dz_{\text{A}}^{\mu
\nu}\right)  
%\nonumber \\ && \quad \times
\exp\left[  -\sum_{\mu,\nu}z_{\text{A}}^{\mu\nu}\left(  NQ_{\text{A}}^{\mu\nu
}-\sum_{i=1}^{N}{\cal A}_i^{\mu\nu}\right)  \right]  =1,
\end{eqnarray}%
\begin{eqnarray}
\int_{-\infty}^{\,\infty}\prod_{\mu,\nu}\left(  dQ_{\text{B}}^{\mu\nu}\right)
\int_{-i\infty}^{i\infty}\prod_{\mu,\nu}\left(  \frac{N}{2\pi i}dz_{\text{B}}^{\mu
\nu}\right) 
%\nonumber \\ && \quad \times
 \exp\left[  -\sum_{\mu,\nu}z_{\text{B}}^{\mu\nu}\left(  NQ_{\text{B}}^{\mu\nu
}-\sum_{i=1}^{N}{\cal B}_i^{\mu\nu}\right)  \right]  =1,
\end{eqnarray}
where we have used the complex representation of the Dirac $\delta$-function.
We then write
\begin{eqnarray}
&& Z  = \int[dz]\left[  \sum_{\{\bm{a}_{i}\}}\exp\left(  \sum_{i}\sum_{\mu,\nu
}z_{\text{A}}^{\mu\nu}{\cal A}_i^{\mu\nu}\right)  \right]  \left[  \sum_{\{\bm{b}_{i}%
\}}\exp\left(  \sum_{i}\sum_{\mu,\nu}z_{\text{B}}^{\mu\nu}{\cal B}_i^{\mu\nu}\right)
\right]  \int[dQ]
\nonumber\\
&&  \cdot
\exp\left\{  \frac{\beta N}{2}\sum_{\mu,\nu}\left[  \varepsilon_{\text{A}}{Q_{\text{A}}%
^{\mu\nu}}^{2}+\varepsilon_{\text{B}}{Q_{\text{B}}^{\mu\nu}}^{2}+2JQ_{\text{A}}^{\mu\nu}Q_{\text{B}}^{\mu\nu}-\frac
{2}{\beta}\left(  z_{\text{A}}^{\mu\nu}Q_{\text{A}}^{\mu\nu}+z_{\text{B}}^{\mu\nu}Q_{\text{B}}^{\mu\nu
}\right)  \right]  \right\}  ,
\end{eqnarray}
where $[dz]=\prod_{\mu,\nu}dz_{\text{A}}^{\mu\nu}dz_{\text{B}}^{\mu\nu}$, $[dQ]=\prod
_{\mu,\nu}dQ_{\text{A}}^{\mu\nu}dQ_{\text{B}}^{\mu\nu}$, and we have discarded contributions
to the free energy of $\mathcal{O}(\ln N)$. Since the interaction energies,
given by Eqs. (\ref{ea-eq}-\ref{eab-eq}), are of infinite range, the sum over
states has been reduced to a simple problem, with decoupled sites, and the
final results are of mean-field nature. The sum over the A variables leads
to%
\begin{eqnarray}
&& \sum_{\{\bm{a}_{i}\}}\exp\left(  \sum_{i}\sum_{\mu,\nu}z_{\text{A}}^{\mu\nu
}{\cal A}_i^{\mu\nu}\right) =\left[  \sum_{\{\bm{a}\}}\exp\left(
\sum_{\mu,\nu}z_{\text{A}}^{\mu\nu}{\cal A}^{\mu\nu}\right)  \right]  ^{N}
\nonumber \\ &&
\quad = \left\{  \exp\left[  \ln2-\frac{1}{2}\sum_{\nu}z_{\text{A}}^{\nu\nu}+\ln\left(
\sum_{\nu}e^{3z_{\text{A}}^{\nu\nu}/2}\right)  \right]  \right\}^{N},
\label{tracea-eq}%
\end{eqnarray}
which also holds with the exchange A$\leftrightarrow$B.

For large $N$, the integrals over the $Q$ variables can be obtained by the
Laplace method. To leading order, we have%
\begin{eqnarray}
I_{Q}=\exp\left[  -\frac{N}{2\beta\left(  \varepsilon_{\text{A}}\varepsilon_{\text{B}}-J^{2}\right)  }\sum_{\mu,\nu
}\left(  \varepsilon_{\text{B}}{z_{\text{A}}^{\mu\nu}}^{2}+\varepsilon_{\text{A}}{z_{\text{B}}^{\mu\nu}}^{2}
%\right. \right. \nonumber \\ && \quad - \left. \left.
- 2Jz_{\text{A}}^{\mu\nu}%
z_{\text{B}}^{\mu\nu}\right)  \right].
\label{iq-eq}%
\end{eqnarray}
The stationary point conditions,
\begin{eqnarray}
&& z_{\text{A}}^{\mu\nu}=\beta\left(  \varepsilon_{\text{A}}Q_{\text{A}}^{\mu\nu}+JQ_{\text{B}}^{\mu\nu}\right),
\nonumber \\  &&
z_{\text{B}}^{\mu\nu}=\beta\left(  \varepsilon_{\text{B}}Q_{\text{B}}^{\mu\nu}+JQ_{\text{A}}^{\mu\nu}\right),
\label{ztoq-eq}%
\end{eqnarray}
which can also be written as%
\begin{equation}
Q_{\text{A}}^{\mu\nu}=\frac{\left(  \varepsilon_{\text{B}}z_{\text{A}}^{\mu\nu}-Jz_{\text{B}}^{\mu\nu}\right)  }%
{\beta\left(  \varepsilon_{\text{A}}\varepsilon_{\text{B}}-J^{2}\right)  },\quad Q_{\text{B}}^{\mu\nu}=\frac{\left(
\varepsilon_{\text{A}}z_{\text{B}}^{\mu\nu}-Jz_{\text{A}}^{\mu\nu}\right)  }{\beta\left(  \varepsilon_{\text{A}}\varepsilon_{\text{B}}-J^{2}\right)},
\label{qtoz-eq}%
\end{equation}
may be used to switch to the former $Q$ variables. Eqs. \eqref{tracea-eq} and
\eqref{iq-eq} can be used to write the partition function
\begin{eqnarray}
Z &=& e^{2N\ln2}\int[dz]\exp\left\{  -N\left[  \frac{1}{2\beta\left(
\varepsilon_{\text{A}}\varepsilon_{\text{B}}-J^{2}\right)  }\sum_{\mu,\nu}\left(  \varepsilon_{\text{B}}{z_{\text{A}}^{\mu\nu}}^{2}+\varepsilon_{\text{A}}{z_{\text{B}}^{\mu\nu
}}^{2}-2Jz_{\text{A}}^{\mu\nu}z_{\text{B}}^{\mu\nu}\right)  
\right. \right. \nonumber \\ && \quad + \left. \left.
 \frac{1}{2}\sum_{\nu}\left(  z_{\text{A}}^{\nu\nu}+z_{\text{B}}%
^{\nu\nu}\right)  -\ln\left(  \sum_{\nu}e^{3z_{\text{A}}^{\nu\nu}/2}\right)
-\ln\left(  \sum_{\nu}e^{3z_{\text{B}}^{\nu\nu}/2}\right)  \right]  \right\}  .
\end{eqnarray}
Note that, on the grounds of mathematical simplicity, we decided to begin
\ the integrations by the $Q$ rather than the $z$ variables.

The complex integrals over the $z$ variables may be done by the method of
steepest descents. For $\mu\neq\nu$, the intergral over $z_{\mu\nu}$ gives a
contribution of $\mathcal{O}(\ln N)$, which will then be discarded. The
saddle-point conditions for the diagonal variables lead to the self-consistent
equations
\begin{eqnarray}
Q_{\text{A}}^{\mu\mu}=\frac{1}{2}\left[  3\frac{e^{3\beta\left(  \varepsilon_{\text{A}}Q_{\text{A}}^{\mu\mu
}+JQ_{\text{B}}^{\mu\mu}\right)  /2}}{\sum_{\nu}e^{3\beta\left(  \varepsilon_{\text{A}}Q_{\text{A}}^{\nu\nu
}+JQ_{\text{B}}^{\nu\nu}\right)  /2}}-1\right]  ,
\end{eqnarray}%
\begin{eqnarray}
Q_{\text{B}}^{\mu\mu}=\frac{1}{2}\left[  3\frac{e^{3\beta\left(  \varepsilon_{\text{B}}Q_{\text{B}}^{\mu\mu
}+JQ_{\text{A}}^{\mu\mu}\right)  /2}}{\sum_{\nu}e^{3\beta\left(  \varepsilon_{\text{B}}Q_{\text{B}}^{\nu\nu
}+JQ_{\text{A}}^{\nu\nu}\right)  /2}}-1\right]  ,
\end{eqnarray}
where we have switched to the $Q$ variables using Eqs. \eqref{ztoq-eq} and
\eqref{qtoz-eq}. Note that we recover the simple MSZ model in the limit
$J\rightarrow0$, and that the $\mathbb{Q}$ tensors are traceless. The free
energy is finally written as%
\begin{eqnarray}
&& f = -2kT\ln2+\frac{1}{2}\sum_{\nu}\left(  \varepsilon_{\text{A}}{Q_{\text{A}}^{\nu\nu}}%
^{2}+\varepsilon_{\text{B}}{Q_{\text{B}}^{\nu\nu}}^{2}+2JQ_{\text{A}}^{\nu\nu}Q_{\text{B}}^{\nu\nu}\right) - k T
\nonumber \\ &&
\cdot \left\{  \ln\left[  \sum_{\nu} \exp \left( \frac{3}{2} \, \beta\left(
\varepsilon_{\text{A}}Q_{\text{A}}^{\nu\nu}
+JQ_{\text{B}}^{\nu\nu}\right) \right) \right] 
+ \ln \left[  \sum_{\nu
} \exp \left( \frac{3}{2} \, \beta \left(  \varepsilon_{\text{B}}Q_{\text{B}}^{\nu\nu}+JQ_{\text{A}}^{\nu\nu}\right)  \right) \right]
\right\}  .
\end{eqnarray}

\section{Global phase diagram}

The equations of state and the free energy are further simplified if we
consider the standard parametric form of the order parameter tensors,
\begin{equation}
\mathbb{Q}_{\text{A}}\equiv\left(
\begin{array}
[c]{ccc}%
-\frac{1}{2}S_{\text{A}} & 0 & 0\\
0 & -\frac{1}{2}S_{\text{A}} & 0\\
0 & 0 & S_{\text{A}}%
\end{array}
\right)  ,
\end{equation}%
\begin{equation}
\mathbb{Q}_{\text{B}}\equiv\left(
\begin{array}
[c]{ccc}%
-\frac{1}{2}S_{\text{B}} & 0 & 0\\
0 & -\frac{1}{2}S_{\text{B}} & 0\\
0 & 0 & S_{\text{B}}%
\end{array}
\right)  ,
\end{equation}
so that
\begin{eqnarray}
&& S_{\text{A}}=\frac{1}{2}\left[  \frac{3}{1+2\,e^{-9(S_{\text{A}}+JS_{\text{B}})/4T}}-1\right],
\nonumber \\ &&
S_{\text{B}}=\frac{1}{2}\left[  \frac{3}{1+2\,e^{-9(\varepsilon_{\text{B}}S_{\text{B}}+JS_{\text{A}})/4T}%
}-1\right]  ,
\label{sasb-final}
\end{eqnarray}
and
\begin{eqnarray}
f  &=& -2T\ln2+\frac{3}{4}\left[  S_{\text{A}}(S_{\text{A}}-2)+\varepsilon_{\text{B}}S_{\text{B}}(S_{\text{B}}-2) +2J
%\right. \nonumber \\ &&  \times \left.
(S_{\text{A}} S_{\text{B}} - (S_{\text{A}}+S_{\text{B}}))\right] 
\nonumber \\ && \quad 
-T\ln\left[  1+2\,e^{-9(S_{\text{A}}+JS_{\text{B}})/4T}\right]  
%\nonumber \\ && \times
\left[
1+2\,e^{-9(\varepsilon_{\text{B}}S_{\text{B}}+JS_{\text{A}})/4T}\right].
\label{fen-eq-final}%
\end{eqnarray}
Note that we have written the free energy $f$ and the energy parameters $\varepsilon_{\text{B}}$
and $J$, as well as the temperature $T$, in terms of the energy parameter $\varepsilon_{\text{A}}$.

\begin{figure}[th]
\vspace{0.5cm}
\par
\begin{center}
\includegraphics[width=0.6\linewidth]{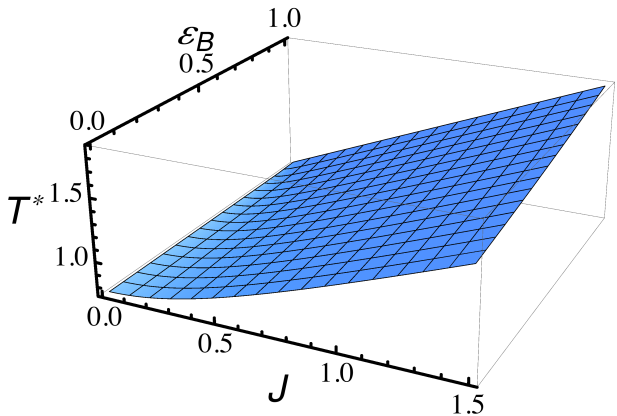}
\end{center}
\caption{Stability threshold of the isotropic phase in terms of the spinodal temperature $T^*$ and the energy
parameters $\varepsilon_{\text{B}}$ and $J$. The isotropic phase is unstable below the surface.}%
\label{stability}%
\end{figure}

Assuming $\varepsilon_{\text{B}}<\varepsilon_{\text{A}}$, we can write an expansion of the free energy $f$ in
terms of $S_{\text{A}}$, up to $\mathcal{O} ({S_{\text{A}}}^{2})$,%
\begin{eqnarray}
f= -2\,T\ln6+\frac{3}{16}\frac{\left(  3\varepsilon_{\text{B}}-3J^{2}-4T\right)  }{\left(
3\varepsilon_{\text{B}}-4T\right)  ^{2}T}
%\nonumber \\ && 
 \left[  9J^{2}-(3-4T)(3\varepsilon_{\text{B}}-4T)\right]  {S_{\text{A}}}^{2},
\label{f-exp-landau}
\end{eqnarray}
so that the coefficient of ${S_{\text{A}}}^2$ is proportional to $T-T^*$, where $T^*$ is given by
\begin{eqnarray}
T^* = \frac{3}{8} \left(1+\varepsilon_{\text{B}} + \sqrt{(1+\varepsilon_{\text{B}})^2+4(J^2-\varepsilon_{\text{B}})} \right).
\end{eqnarray}
If we suppose that the coefficients of higher powers of $S_{\text{A}}$ in \eqref{f-exp-landau} are weakly dependent on temperature, $T^*$ is associated with the spinodal temperature of a Landau-de Gennes theory of nematics, and establishes the threshold of stability of the isotropic phase. For $\varepsilon_{\text{B}}\rightarrow0$, the spinodal temperature is given by $T^*=1+\sqrt{1+J^{2}}$, which increases with the coupling parameter $J$, and is thus an indication of
the enhancement of the nematic phase. For $\varepsilon_{\text{B}}\rightarrow1$, $T^*$ increases
approximately linearly with $J$. The global behavior is shown in the diagram of Fig. \ref{stability}, in terms of $\varepsilon_{\text{B}}$, $J$, and $T^*$. In particular, the coefficient of ${S_{\text{A}}}^{2}$ becomes proportional to $(T-3/4)$ in the limit of zero coupling between the subsystems, $J\rightarrow0$, in agreement with previous results for the simple Maier-Saupe-Zwanzig model \cite{carmo10}.

\begin{figure}[th]
\vspace{0.5cm}
\par
\begin{center}
\includegraphics[width=0.6\linewidth]{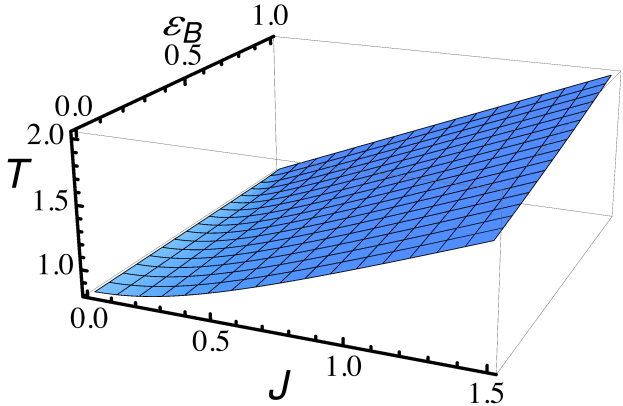}
\end{center}
\caption{Global phase diagram in terms of temperature $T$ and energy
parameters $\varepsilon_{\text{B}}$ and $J$. The transition from the nematic (below the surface) to
the isotropic phase (above the surface) is discontinuous, with a jump in the
nematic order parameters $S_{\text{A}}$ and $S_{\text{B}}$.}%
\label{phase-diagram}%
\end{figure}

In Fig \ref{phase-diagram}, we show the global phase diagram of this model in
terms of the temperature $T$ and the energy parameters $\varepsilon_{\text{B}}$ and $J$, obtained from numerical solutions of the exact equations of state \eqref{sasb-final} and free energy \eqref{fen-eq-final}. The
transition from the nematic (low-temperature) to the isotropic
(high-temperature) phase is of first order, with a jump in both nematic order
parameters $S_{\text{A}}$ and $S_{\text{B}}$. The nematic phase is enhanced by the increase
of the energy parameters $\varepsilon_{\text{B}}$ and $J$. As it should be anticipated, this surface is very similar to the threshold of stability of the isotropic phase.

\section{Suspensions of ferroelectric nanoparticles in nematic systems}

A coupled system of two types of nematic subsystems has been considered by
Lopatina and Selinger to represent a dilute suspension of ferroelectric
nanoparticles in a nematic host \cite{lopatina09,lopatina11}. The mechanism
behind this mapping is the effect on the nematic mesogenic units of the dipole
aligning electric fields produced by the ferroelectric nanoparticles. For a
range of parameters, assuming uniform nematic order, one may integrate out the
position variables, and thus eliminate the complicated spatial dependence of
the interactions between mesogens and nanoparticles. It is then possible to
associate a nematic-like order parameter with the distribution of orientations of the dipole moments
of the nanoparticles. The energy of interaction turns out to be proportional
to $S_{\text{LC}}\cdot S_{\text{NP}}$, where $S_{\text{LC}}$ and
$S_{\text{NP}}$ represent the nematic order parameters of the liquid crystal
and nanoparticle systems respectively \cite{lopatina09,lopatina11}, and which has essentially the same form
as the coupling energy of Eq. \eqref{eab-eq}. Though the interaction among the
nanoparticles has dipole symmetry, one expects the limit $\varepsilon_{\text{B}}\rightarrow0$ will
approximately represent a situation in the dilute regime.

\begin{figure*}[th]
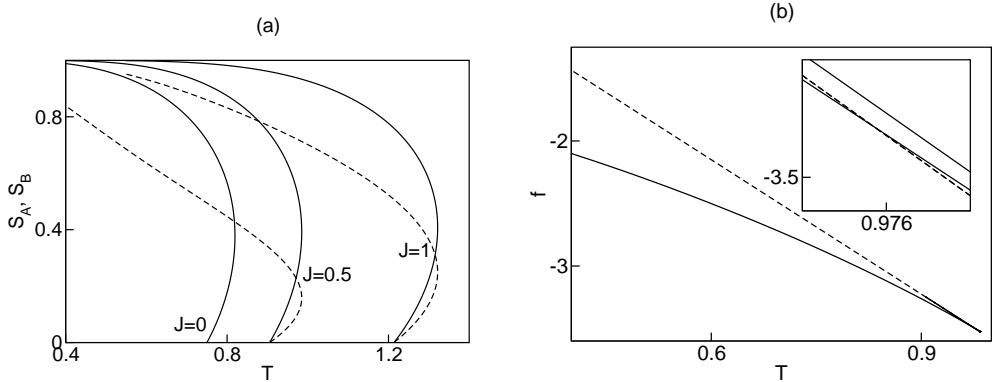

\vspace{0.5cm} \begin{minipage}[b]{0.45\linewidth}
\includegraphics[width=\linewidth]{fig3a.eps}
\end{minipage}
\hspace{0.5cm} \begin{minipage}[b]{0.45\linewidth}
\includegraphics[width=\linewidth]{fig3b.eps}
\end{minipage}
\caption{(a) Nematic order solutions for $S_{\text{A}}$ (full curves) and $S_{\text{B}}$
(dashed curve) as a function of temperature for $\varepsilon_{\text{B}}=0$, and some values of $J$.
There may be up to three solutions: stable nematic, unstable nematic,
and isotropic ($S=0$). The transition from the nematic (upper solution) to
the isotropic phase ($S=0$) takes place where the two free energies become
equal. (b) Free energy as a function of temperature for the stable and
unstable nematic (full curves) and the isotropic solution (dashed curve), with
$\varepsilon_{\text{B}}=0$ and $J=0.5$. The inset shows a magnification of the region where the stable
nematic and the isotropic solutions have the same free energy.}%
\label{order-fen}%
\end{figure*}

In Fig \ref{order-fen}a we show the nematic solutions for $S_{\text{A}}$ (full
curves) and $S_{\text{B}}$ (dashed curves) as a function of temperature, for $\varepsilon_{\text{B}}=0$
and some values of the coupling parameter $J$. In general, there may be up to
three solutions, two nematic and one isotropic. The lower curve is unstable,
and the nematic-isotropic transition takes place at the temperature at which
the stable nematic (upper curve) and isotropic ($S=0$) free energies become
equal. The free energies for the three solutions are shown in Fig
\ref{order-fen}b, for $J=0.5$. In agreement with some previous experimental
\cite{li06} and theoretical studies \cite{lopatina09,lopatina11,gorkunov11},
the nematic order is enhanced by the coupling with an ordered subsystem B,
which may be interpreted as representing the set of local orientations of
ferroelectric nanoparticles. Note that our model always predicts an increase of $T_{\text{NI}}$ with the addition of ferroelectric nanoparticles, even though a scenario where the nematic-isotropic transition temperature decreases has also been observed \cite{kurochkin10}. This scenario is probably suppressed by the assumption of uniform nematic order around the nanoparticles, which our model inherits from Lopatina and Selinger's derivation of the nematic-nanoparticle coupling \cite{lopatina09,lopatina11}.

\section{Summary}

We have used a Maier-Saupe-Zwanzig lattice model to study a coupled system of
two types of nematic subsystems. The model is simple enough to allow an exact
calculation of the free energy. In contrast to the usual phenomenological
approaches, it explicitly contains the microscopic energy parameters involved
in the nematic-isotropic transition. We show that the nematic-isotropic
transition temperature of subsystem A increases with both the interaction
energy among mesogens of system B and the two-subsystem coupling $J$. We
draw the global phase diagram in terms of the temperature $T$ and the energy
parameters. For non-interacting mesogens in system B, this model may be used to describe the
experimentally observed enhancement of the nematic ordering produced by
ferroelectric nanoparticles suspended in a liquid-crystalline host.\bigskip

\textbf{Acknowledgements:} We thank the financial support of the Brazilian agencies Fapesp and CNPq.

\end{document}